# The Bounds on the magnetic moment of the τ-neutrino via the process $e^+e^- \to \nu\bar{\nu}\gamma$

C.Aydın*, M.Bayar, and N.Kılıç

Karadeniz Technical University Department of Physics, 61080 Trabzon, Turkey

*coskun@ktu.edu.tr

**Abstract**

Bounds on the anamolous magnetic moment of the τ neutrino are calculated through the reaction $e^+e^- \to \nu\bar{\nu}\gamma$ at the neutral boson pole and in the framework of an extended standard model, a left-right symmetric model and a superstring-inspired $E_6$ model which has one extra low-energy neutral gauge boson. The results are based on the recent data reported by the L3 Collaboration at CERN LEP.

## I. INTRODUCTION

The question of whether the neutrinos are Dirac or Majorana particles is one of the most important issue in particle physics, astrophysics and cosmology. As known that the properties of neutrinos have become the subject of an increasing research effort over the last years. The search for the neutrino mass, magnetic moment, dipole moment and anapole moment is of great significance for the choose of theory of elementary particles and for understanding of phenomena such a supernova dynamics , stellar evolution and the production of neutrino by the sun [1].

The neutrinos called Weyl neutrinos are massless in the standard model (SM). In many extensions of the Standard Model the neutrino acquires a nonzero mass. Massive neutrinos are Dirac or Majorana neutrinos. These neutrinos have different electromagnetic properties. Dirac neutrino has three form factors which are charge, magnetic moment and anapole moment since the electric dipole moment is zero in CP conserving theory [2]. Majorana neutrino has only one form factor which is the anapole moment [3]. In this manner the neutrinos seem to be likely candidates for carrying features of physics beyond the Standard Model. Apart from masses and mixings, magnetic moments, electric dipole moments and anapole moments are also signs of new physics. These were calculated many authors in different model [4].

In 1994, Gould and Rothstein [5] reported a bound on the tau neutrino magnetic moment which they obtained through the analysis of the process $e^+e^- \to \nu\bar{\nu}\gamma$, near the $Z_0$-resonance by considering a massive tau neutrino and using standard model $Ze^+e^-$ and $Z\nu\bar{\nu}$ couplings. At low center of mass energy $s << M_{Z_0}^2$, the dominant contribution of this process involves the exchange of virtual photon [6]. The dependence on the magnetic moment comes to a direct coupling to the virtual photon and the observed photon is a result of initial state bremsstrahlung.

In 2001, Aydemir and Sever [7] calculated the same process in framework of a class of $E_6$ inspired models with a light additional neutral vector boson. They obtained for $K_{\nu_\tau} <$ the magnetic moment of the tau neutrino $1.83 \times 10^{-6} \mu_B$. Detailed discussion on $E_6$ inspired model can be found in the literature [8].

Recently, a bound on the tau neutrino magnetic moment (and the tau neutrino dipole moment) has been reported by Gutierrez-Rodriguez, Hermander-Ruiz and Del Rio-De Santiago [9] through the analysis of the process $e^+e^- \to \nu\bar{\nu}\gamma$ in the framework of the left-right symmetric model, based on the $SU(2)_R \times SU(2)_L \times U(1)$ gauge group. Detailed discussion on LRSM can be found in literature [10]. They did their analysis near the resonance of the $Z_1$ $\left(s = M_{Z_1}^2\right)$. Thus their results are independent of the mass of the additional heavy $Z_2$ gauge boson which appears in these kinds of models. Therefore, they have the mixing angle ϕ between the left-right bosons as the only additional parameters besides the SM parameters.

Upper limits on the tau neutrino magnetic moment reported in the literature are $\mu_{\nu_\tau} \leq 3.3 \times 10^{-6} \mu_B$ (90 % C.L.) from a sample of $e^+e^-$ annihilation events collected with the L3 dedector [11] at the $Z_1$ resonance corresponding to an integrated luminosity of 137 $pb^{-1}$; $\mu_{\nu_\tau} \leq 2.7 \times 10^{-6} \mu_B$ (95 % C.L) at $q^2 = M_{z1}^2$ from measurements of the $Z_1$ invisible width at LEP[11] ; $\mu_{\nu_\tau} \leq 1.83 \times 10^{-6} \mu_B$ (90 % C.L.) from the analysis of $e^+e^- \to \nu\bar{\nu}\gamma$ in a class of $E_6$ inspired model [7]; from the order of $\mu_{\nu_\tau} < O\left(1.1 \times 10^{-6} \mu_B\right)$ Akama et al. [12] derive and apply model independent bounds on the anomolous magnetic moment.

As known that there are a number of possible physical processes involving a neutrino with a magnetic moment. Among there are the $\nu - e$ scattering, spin-flavor precession in an external magnetic field, plasmon decay, and the neutrino decay. Our aim in this study is to analyse the reaction $e^+e^- \to \nu\bar{\nu}\gamma$ in the different model (beyond SM). We use recent data collected with the L3 dedector at CERN $e^+e^-$ collider LEP [11] near the vector boson resonance in framework of beyond SM and we attribute an anomalous magnetic moment to a massive tau neutrino.

At higher s, near the Z pole $s = M_Z^2$, the dominant contribution for $E_\gamma > 10$ GeV involves the exchange of a Z boson. The dependence on the magnetic moment now comes from the radiation of the photon observed by the neutrino an antineutrino in the final. The Feynman diagrams which give the most important contribution to the cross section are shown in Fig.1.

We emphasize here the importance of the final state radiation near the Z pole, which occurs preferentially at high E compared to conventional bremsstrahlung.

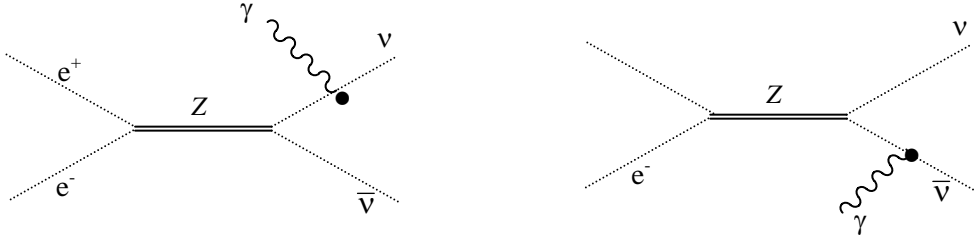

**Figure 1.** The Feynman diagrams contributing to the process $e^+e^- \to \nu\bar{\nu}\gamma$.

We calculate the total cross section of the process $e^+e^- \to \nu\bar{\nu}\gamma$ using the Breit-Wigner resonance form [13]

$$\sigma = \frac{12\pi \Gamma_{z \to e^+e^-} \Gamma_{z \to \nu\bar{\nu}\gamma}}{\left(s - M_Z^2\right)^2 + M_Z^2 \Gamma_Z^2} \tag{1}$$

where $\Gamma_{Z \to e^+e^-}$ in the decay rate of Z to the channel $Z \to e^+e^-$, $\Gamma_{Z \to \nu\bar{\nu}\gamma}$ in the decay rate Z to the channel $Z \to \nu\bar{\nu}\gamma$ and $\Gamma_Z$ is total width of Z ($Z_0$, $Z_l$, $Z_\theta$).

This paper is organized as follows: In Sec.II, . we present the calculations of the process $e^+e^- \to \nu\bar{\nu}\gamma$ in framework of an extended standard model, a left-right symmetric model and a superstring-inspired $E_6$ model, respectively. In SecIII we make the numerical computation and summarize our results.

## II. THE TOTAL CROSS SECTION

a) Width of section $Z \to e^+ e^-$

In this section we calculate the total width of the reaction $Z \to e^+ e^-$ in the context of the extended SM, left-right symmetric model and a superstring-inspired $E_6$ model. The expression for the Feynman amplitude M of the process $Z \to e^+ e^-$ is given by, respectively

$$M = -\frac{ig}{c_w} \bar{u} \gamma^\mu \frac{1}{2}(g_V^e - g_A^e \gamma^5) v \varepsilon_\mu^\lambda \quad \text{for } Z_0 \tag{2}$$

$$M = -\frac{ig}{c_w} \bar{u} \gamma^\mu \frac{1}{2}(a g_V^e - b g_A^e \gamma_5) v \varepsilon_\mu^\lambda \quad \text{for } Z_1 \tag{3}$$

where $g_V^e = -\frac{1}{2} + 2 x_W$, $g_A^e = -\frac{1}{2}$, $x_W = \sin^2 \theta_W$, $a = c_\phi - \frac{s_\phi}{r_w}$, $b = c_\phi + r_w s_\phi$, $s_\varphi = \sin \phi$, $c_\phi = \cos \phi$, $\phi$ is the mixing parameter of the LRSM, $r_W = \sqrt{\cos 2 \theta_W}$, $\theta_W$ is the electroweak mixing angle, $\varepsilon_\mu^\lambda$ is the polarization vector of the boson Z.

$$M = -\frac{ig}{c_w} \bar{u} \gamma^\mu (C'_V - C'_A \gamma_5) v \varepsilon_\mu^\lambda \quad \text{for } Z_\theta \tag{4}$$

where $C'_V = X^{1/2} \left( \frac{1}{\sqrt{6}} \cos \varphi + \frac{1}{\sqrt{10}} \sin \varphi \right)$, $C'_A = 2 X^{1/2} \frac{\sin \varphi}{\sqrt{10}}$, $X = \frac{g_\theta^2}{g^2 + g'^2} \left( \frac{M_{z_0}}{M_{Z_\theta}} \right)^2$, $\varphi$ is mixing angle $Z_\psi$ and $Z_\chi$ [8].

The expression for the total width of the process $Z \to e^+ e^-$, due only to the Z boson exchange, according to the diagrams depicted in Fig.1, and using the expression for the amplitude given in Eq.2(3,4), is respectively

$$\Gamma_{Z_0 \to e^+ e^-} = \frac{\alpha (1 - 4 x_W + 8 x_W^2)}{24 \pi x_W (1 - x_W)} M_{z_0} \tag{5}$$

$$\Gamma_{Z_1 \to e^+ e^-} = \frac{\alpha M_{Z_1}}{24 x_W (1 - x_W)} \left[ \frac{1}{2}(a^2 + b^2) - 4 a^2 x_W + 8 a^2 x_W^2 \right] \tag{6}$$

$$\Gamma_{Z_\theta \to e^+ e^-} = \frac{\alpha M_{z_\theta}}{12 x_W (1 - x_W)} (C_V'^2 + C_A'^2) \tag{7}$$

b) Width of $Z \to \nu\bar{\nu}\gamma$

The expression for the Feynman amplitude of the process $Z \to \nu\bar{\nu}\gamma$ is due only to the $Z$ boson exchange, as shown in the diagrams in Fig.1.

In the LRSM, the expression for the Feynman amplitude M of the process $Z \to \nu\bar{\nu}\gamma$ is given by

$$M = \frac{ig}{4\cos\theta_w} \kappa k^\nu \varepsilon^\mu_\gamma \bar{u}(q_1) \left[ \frac{1}{(k+q_1)^2 - m_\nu^2} \sigma_{\mu\nu} (\not{k} + \not{q}_1 + m_\nu) \not{\varepsilon} (a - b\gamma_5) \right.$$
$$\left. + \frac{1}{(k+q_2)^2 - m_\nu^2} \not{\varepsilon}(a - b\gamma_5)(\not{k} + \not{q}_2 + m_\nu)\sigma_{\mu\nu} \right] v(q_2) \quad \text{for } Z_1 \quad (8)$$

where $k$ is the photon momentum, $q_1$ is the neutrino momentum, $q_2$ is the antineutrino momentum, $\varepsilon^\lambda_\gamma$ and $\varepsilon^\lambda_Z$ are the polarization vectors of photon and of the boson $Z$, respectively. We obtain the matrix element of $Z_0$ for $a = b = 1$ and $Z_\theta$ for $a = b = X^{1/2}\left(-\frac{1}{\sqrt{6}}\cos\varphi + \frac{3}{\sqrt{10}}\sin\varphi\right)$ in Eq.(8).

After long and straightforward calculation, we obtain for the total width of $Z \to \nu\bar{\nu}\gamma$

$$\Gamma_{Z_0 \to \nu\bar{\nu}\gamma} = \int\int_{E_\gamma \theta} \frac{\alpha\kappa^2}{96\pi^2 x_W(1-x_W)M_{Z_0}} \left[2\left(s - 2\sqrt{s}E_\gamma\right) + \frac{2}{3}E_\gamma^2\right] E_\gamma dE_\gamma \sin\theta \, d\theta \quad (9)$$

$$\Gamma_{Z_1 \to \nu\bar{\nu}\gamma} = \int\int_{E_\gamma \theta} \frac{\alpha\kappa^2}{96\pi^2 x_W(1-x_W)M_{Z_1}} \left[(a^2+b^2)\left(s - 2\sqrt{s}E_\gamma\right) + \frac{2a^2}{3}E_\gamma^2\right] E_\gamma dE_\gamma \sin\theta \, d\theta \quad (10)$$

$$\Gamma_{Z_\theta \to \nu\bar{\nu}\gamma} = \int\int_{E_\gamma \theta} \frac{\alpha\kappa^2 a^2}{96\pi^2 x_W(1-x_W)M_{Z_\theta}} \left[2\left(s - 2\sqrt{s}E_\gamma\right) + \frac{2}{3}E_\gamma^2\right] E_\gamma dE_\gamma \sin\theta \, d\theta \quad (11)$$

where $E_\gamma$ and $\cos\theta_\gamma$ are the energy and scattering angle of the photon.

The substitution of Eq. 5( 6,7) and 9( 10,11) in Eq.(1) gives, respectively

$$\sigma = \frac{\alpha^2 \kappa^2 (1 - 4x_W + 8x_W^2)}{192\pi^2 x_W^2 (1-x_W)^2 M_{Z_0} \Gamma^2_{M_{Z_0}}} \int\int_{E_\gamma \theta} \left[2\left(s - 2\sqrt{s}E_\gamma\right) + \frac{2}{3}E_\gamma^2\right] E_\gamma dE_\gamma \sin\theta \, d\theta \quad (12)$$

$$\sigma = \frac{\alpha^2 \kappa^2}{192\pi^2 x_W^2 (1-x_W)^2 M_{Z_1} \Gamma_{M_{Z_1}}^2} \left[ \frac{1}{2}(a^2+b^2) - 4a^2 x_W + 8a^2 x_W^2 \right]$$
$$\times \iint_{E_\gamma \theta} \left[ (a^2+b^2)(s - 2\sqrt{s} E_\gamma) + \frac{2a^2}{3} E_\gamma^2 \right] E_\gamma dE_\gamma \sin\theta \, d\theta \quad (13)$$

$$\sigma = \frac{\alpha^2 \kappa^2}{96\pi^2 x_W^2 (1-x_W)^2 M_{Z_\theta} \Gamma_{Z_\theta}^2} (C_V^2 + C_A^2)$$
$$\times \iint_{E_\gamma \theta} \left[ 2(s - 2\sqrt{s} E_\gamma) + \frac{2}{3} E_\gamma^2 \right] E_\gamma dE_\gamma \sin\theta \, d\theta \quad (14)$$

where $\theta$ from 44.5° to 135.5° and $E_\gamma$ from 15 GeV to $\sqrt{s}/2$,

$$C_V = X \left( \frac{1}{\sqrt{6}} \cos\varphi + \frac{1}{\sqrt{10}} \sin\varphi \right) \left( -\frac{1}{\sqrt{6}} \cos\varphi + \frac{3}{\sqrt{10}} \sin\varphi \right)$$

$$C_A = 2X \frac{\sin\varphi}{\sqrt{10}} \left( \frac{-\cos\varphi}{\sqrt{6}} + \frac{3}{\sqrt{10}} \sin\varphi \right).$$

### III. NUMERICAL CALCULATIONS AND DISCUSSION

The L3 Collaboration evaluated the selection efficiency using detector-simulated $e^+e^- \rightarrow \nu\bar{\nu}\gamma$ events. A total of 14 events was found by the selection. As was discussed in Ref.[4] $N = \sigma \cdot L$ be less than 14, with $L = 137\, pb^{-1}$, according to the data reported by the L3 Collabaration Ref.[11]. Using the following numeric values: $x_W = \sin^2\theta_W = 0.2314$, $\Gamma_Z = 2.49\, GeV$, assuming all $\Gamma_Z$ ($Z_0$, $Z_1$, $Z_\theta$) to be the same, $\alpha = e^2/4\pi$ is the fine structure constant. Taking this into consideration, we put bound for the tau neutrino magnetic moment $\kappa = 6.59 \cdot 10^{-6} \mu_B$ in the extended SM. Using above N and L values we have obtained results in Table1 and 2 for the LRSM and superstring-inspired $E_6$ model which is assumed all U(1) couplings to be same, i.e., $\frac{g_\theta^2}{g^2 + g'^2} = \frac{5}{3} \sin^2\theta_W$. We have seen that these results depend on mixing angle, N and L in the LRSM and the superstring-inspired $E_6$ model.

**TABLE 1**. Bounds on the $\mu_{\nu_\tau}$ magnetic moment for different of mixing angle $\phi$ in the $Z_1$ resonance, i.e., s=$\left(M_{Z_1}^2\right)$ (Second column values are calculated by A. Gutierrez-Rodriguez, M. A. Hermander-Ruiz and A. Del Rio-De Santiago).

| $\phi$ | $\kappa(10^{-6}\mu_B)$ | $\kappa(10^{-6}\mu_B)$ |
|---|---|---|
| -0.009 | 3.71 | 3.37 |
| -0.005 |  | 3.34 |
| -0.004 | 3.72 |  |
| 0.000 | 3.71 | 3.31 |
| 0.004 | 3.73 | 3.28 |

**TABLE 2.** Bounds on the $\mu_{\nu_\tau}$ magnetic moment for different of mixing angle $\theta$ and different $M_{Z\theta}$ values.

| $\varphi$ | $\kappa(10^{-6}\mu_B)$ | | | |
|---|---|---|---|---|
|  | $M_{Z_\theta}=91.18\,GeV$ | $M_{Z_\theta}=161\,GeV$ | $M_{Z_\theta}=183\,GeV$ | $M_{Z_\theta}=364.7\,GeV$ |
| 0 | 27.1 | 48.9 | 58 | 55 |
| 37.8° | 39.3 | 70 | 82 | 78 |
| 90° | 6.8 | 12 | 14.4 | 13.7 |
| 127.8° | 9.2 | 16.4 | 19.4 | 18.2 |

We have seen that κ values given in Tab.2 are the bigger than in literature except for some values. . As seen that κ is dependent of L and N values. If the photon energy is greater than half the beam energy, two events are selected from the data [14]. In this case, we obtain the results given in Table 3 in which some of the values are in agreement with the literature for $\varphi= 90°$ and $127.8°$.

**TABLE 3.** Bounds on the $\mu_{\nu_\tau}$ magnetic moment for different of mixing angle $\theta$ and different $M_{Z\theta}$ values for N=2.

| $\varphi$ | $\kappa(10^{-6}\mu_B)$ | | | |
|---|---|---|---|---|
|  | $M_{Z_\theta}=91.18\,GeV$ | $M_{Z_\theta}=161\,GeV$ | $M_{Z_\theta}=183\,GeV$ | $M_{Z_\theta}=364.7\,GeV$ |
| 0 | 10.2 | 18.4 | 22.2 | 20.5 |
| 37.8° | 14.8 | 26.3 | 30.8 | 8 |
| 90° | 2.6 | 4.6 | 5.4 | 5.2 |
| 127.8° | 3.5 | 6.2 | 7.3 | 6.9 |

**Acknowledgements:**

The authors are grateful to Prof.Dr. Rodriguez for fruitful discussions and comments. This work partly supported by the Research Fund of Karadeniz Technical University, under grant contact no 2002.111.001.2.